# Real-time digital signal processor implementation of self-calibrating pulse-shape discriminator for high purity germanium

R. Suarez*, J. L. Orrell, C. E. Aalseth, T. W. Hossbach, and H. S. Miley

Pacific Northwest National Laboratory, 902 Battelle Blvd, Richland, WA 99352, USA

**Abstract**

Pulse-shape analysis of the ionization signals from germanium gamma-ray spectrometers is a method for obtaining information that can characterize an event beyond just the total energy deposited in the crystal. However, as typically employed, this method is data-intensive requiring the digitization, transfer, and recording of electronic signals from the spectrometer. A hardware realization of a real-time digital signal processor for implementing a parametric pulse shape is presented. Specifically, a previously developed method for distinguishing between single-site and multi-site gamma-ray interactions is demonstrated in an on-line digital signal processor, compared with the original off-line pulse-shape analysis routine, and shown to have no significant difference. Reduction of the amount of the recorded information per event is shown to translate into higher duty-cycle data acquisition rates while retaining the benefits of additional event characterization from pulse-shape analysis.

*Corresponding author. Tel.: 509-2335; fax: 509-372-4725

*E-mail address*: **Reynold.Suarez@pnl.gov**

## 1. Introduction

Pulse-shape analysis (PSA) in nuclear gamma-ray spectroscopy is a method of obtaining information beyond just the total ionization charge measured as an indicator of the total energy deposited in the detector. Modern pulse-shape analysis of signals from high-purity germanium (HPGe) photon spectrometers evolved from analog pulse-shape discrimination circuits [1]. The wide availability of digital pulse acquisition systems has allowed HPGe pulse-shape analysis to become dominated by sophisticated algorithmic techniques, typically used in an off-line, post-processing analysis setting. The techniques have primarily pursued either rejection of unwanted event types [2-4] or determining the photon interaction location within an HPGe crystal or segment [5-7]. The techniques themselves are roughly categorized as either parametric methods or fitting methods. The advantage of fitting methods is they, in principle, allow for the greatest extraction of information from each pulse, at the disadvantage of being computationally intensive. In contrast, parametric pulse-shape analysis methods are typically computationally simple and are naturally a form of data reduction, but at the cost of inevitable information loss.

In the current approach we focus on a parametric pulse-shape analysis technique because the advent of waveform digitizing (precision analog-to-digital converters [ADCs]) sets the groundwork for precise and efficient pulse-shape analysis. Digital signal processing technology, designed to perform data filtering and processing efficiently at high speeds, provides the means to implement pulse-shape analysis in real-time. The goal is to provide the benefits of pulse-shape analysis without incurring the overhead required to digitize and record individual pulses [8]. Directly outputting the resulting parameter values rather than the digitized pulse significantly reduces the amount of data recorded for each event. Furthermore, as there is a fixed and limited amount of memory on-board the data acquisition card, decreasing the per event data record size will increase the duty-factor of the acquisition system. Finally, generating the pulse-shape analysis parameters in real-time obviates the need for post-processing of the signal pulses, thus potentially simplifying the analysis phase.

In this work, a semi-coaxial, high-purity, germanium gamma-ray spectrometer was used. An abbreviated description is presented of the parametric pulse-shape analysis method used to discriminate between single-site and multi-site events in such semi-coaxial detectors. Details are presented of the digital signal processor implementation, giving special attention to the translation of an originally off-line C++-based pulse-shape analysis algorithm transformed to meet the requirements of the digital signal processor hardware. The digital signal processor is then used in data collection designed to show the ability of the parametric pulse-shape analysis to preferentially select events having a single-site interaction over those having a multi-site interaction. The collected data is processed both by the digital signal processor and the original off-line pulse-shape analysis algorithm. A comparison of the two shows no difference in the performance of the discrimination of single-site from multi-site events.

## 2.   Pulse-Shape Discrimination

Under the influence of an applied bias voltage, the electron- and hole-charge carriers created by ionizing radiation drift away from the interaction location toward the electrodes. The total charge present in the current signal observed during the drift time is proportional to the amount of energy deposited by ionization. Typically this total charge measurement is the only parameter recorded and used to characterize the radiation interaction process.

To further characterize the radiation interaction process, it is useful to consider the interaction multiplicity. We define multiplicity as the number of ionizing interactions taking place as a consequence of an initial gamma ray entering the germanium crystal. Events are then categorized as low multiplicity (few or single-site interactions) or high multiplicity (multiple interaction locations). Interaction processes characterized by low multiplicity are:  photo-electric absorption of gamma rays less than about 150 keV; double-escape peaks as a result of pair production by gamma rays above about 1 MeV; fast neutron scattering on germanium nuclei; and internal beta decay. High multiplicity events generally occur for gamma-ray energies above about 150 keV where Compton scattering becomes a highly probable gamma-ray interaction mechanism. The pair production process also generates high multiplicity events and begins to contribute above about 1 MeV. Gamma-ray cascade summing (i.e., a cascade sum line) is another example of a high-multiplicity event. Figure 1 illustrates the relative probability of some of these processes as a function of gamma ray energy.

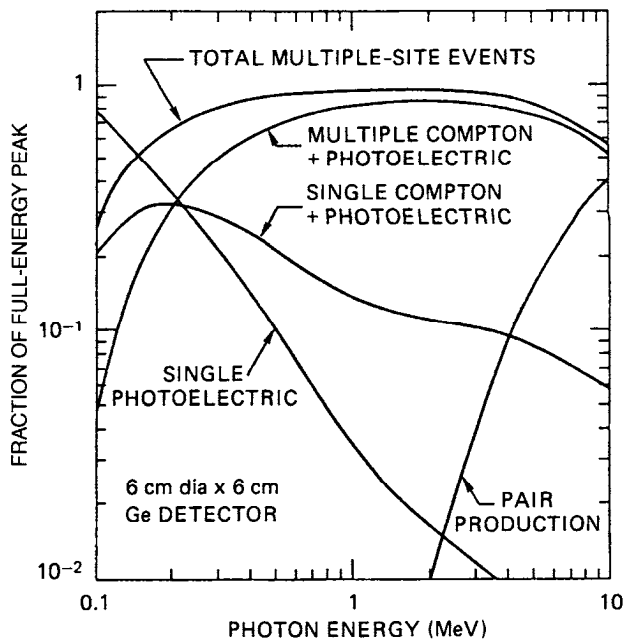

Figure 1: Graph illustrating gamma ray interaction probabilities. From [15].

A previously developed pulse-shape analysis method is used to perform multiplicity-based event discrimination [9]. The technique involves defining a three-dimensional parameter space whose axes correspond to a current pulse's[1] measured width, asymmetry, and normalized moment. These three parameters are simplified calculations bearing resemblance to the moments of a distribution; avoiding implementation of the formal definitions of the skewness and kurtosis was a choice to streamline the DSP algorithms. The pulse width is simply the charge collection

---

[1] The current pulse is obtained by taking a derivative of a voltage pulse from the charge integrating preamplifier.

time – a duration measured in nanoseconds. The equation for the pulse width is illustrated below, where $N_0$ is the window start point, N is the window endpoint, and $\Delta t$ is the pulse sample length.

$$W = (N - N_0)\Delta t$$

The pulse asymmetry provides a measure of the skewness (third moment) of the current pulse. The pulse asymmetry is formulated as the total area (charge) in the front half of the current pulse minus the area of the back half divided by the total pulse area. The equation for the pulse asymmetry is illustrated below, where $F = \sum_{i=N_0}^{N_{\text{mid}}-1} j_i \Delta t$ and $B = \sum_{i=N_{\text{mid}}}^{N} j_i \Delta t$ .

$$\text{Asymmetry} = \frac{F - B}{F + B} \qquad\qquad (1)$$

The pulse asymmetry will vary about 0 with -1 and 1 as the parameter's limiting bounds. An asymmetry value of 0 would correspond to a symmetric pulse. The normalized moment provides a measure of the kurtosis (fourth moment) of the current pulse. The calculation for the normalized moment parameter is defined as:

$$I_n = \frac{\sum_{i=N_0}^{N} j_i \Delta t \left((i - N_{mid})\Delta t\right)^2}{\dfrac{(F + B)W^2}{12}} \qquad\qquad (2)$$

where $N_{\text{mid}}$ is the midpoint of the sampled pulse and $j_i$ is the value of the current at index i. The time granularity of the indices is indicated by $\Delta t$. The numerator in the normalized moment calculation is analogous to a (mass) moment of inertia and the denominator removes (normalizes) the effect of pulses having different widths. One will recognize the factor of 12 in equation 2 from the standard, mass moment of inertia of a rectangular plate. Figure 2 illustrates the calculation of the three pulse-shape parameters.

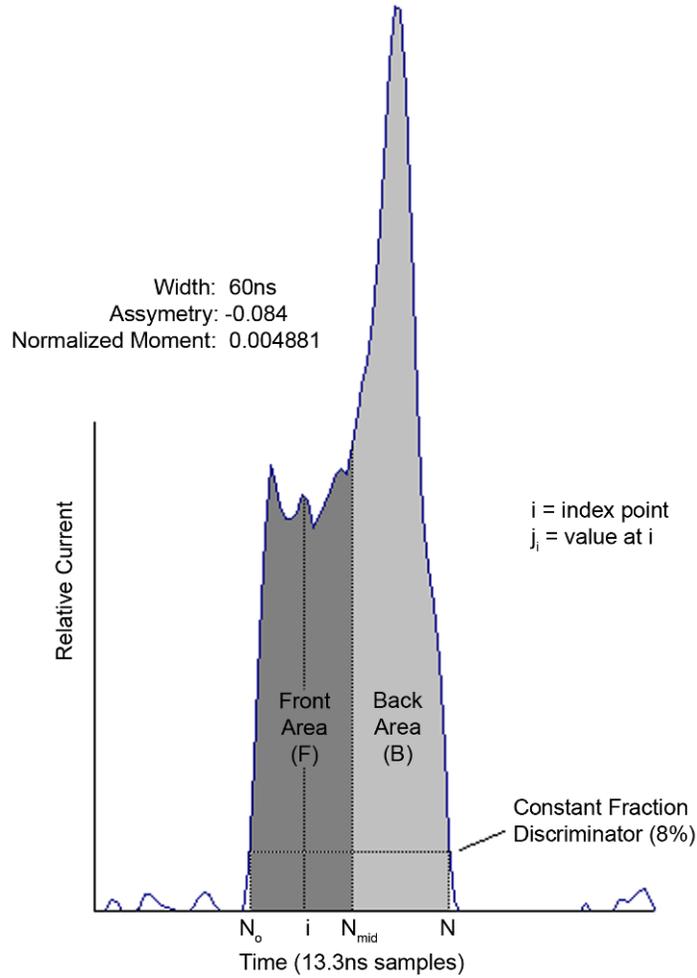

Figure 2: Illustration of the values used in the pulse-shape parameter calculations.

The choice of these parameters is based on the following arguments. For a single-site interaction, the current pulse's width is grossly indicative of the radial position of the interaction [5]. However, a detailed analysis shows the radial interaction position which generates a pulse of minimal width is in fact within the crystal bulk, between the two collection electrodes [10]. The result is the pulse width parameter has values which are degenerate with respect to radial interaction position for single-site interactions. The asymmetry parameter breaks this degeneracy. The approach of charge carriers to the central collection electrode dominates the shape of the current pulse, generally producing the highest current value over the duration of the pulse. Thus for single-site interactions occurring at radii smaller than the radial position of minimal pulse width, the current pulse's maximum will occur 'early' in the pulse. This is because those charge carriers are collected much sooner than the charge carriers traveling toward the outer electrode. In contrast, for single-site interactions occurring at radii lager than the radial position of minimal pulse width, the current pulse's maximum will occur 'late' in the pulse. This is because those charge carriers moving toward the central electrode must travel much further. The asymmetry parameter measures this 'early' versus 'late' collection of charge at the central electrode. Finally, the normalized moment is chosen in consideration of a distinction between single-site and multi-site events. The shape of the current pulse for a single-site event is dominated by the two charge carrier clouds approaching the collection electrodes. Multi-site events, having many more charge carrier clouds, produce a superposed average current pulse having a more uniform height. Consider the simple situation depicted in Figure 3. Both pulse shapes have zero asymmetry; however they will have unequal normalized moments. Thus the normalized moment parameter distinguishes pulses of equal asymmetry, in a way relevant to discrimination between single-site and multi-site events.

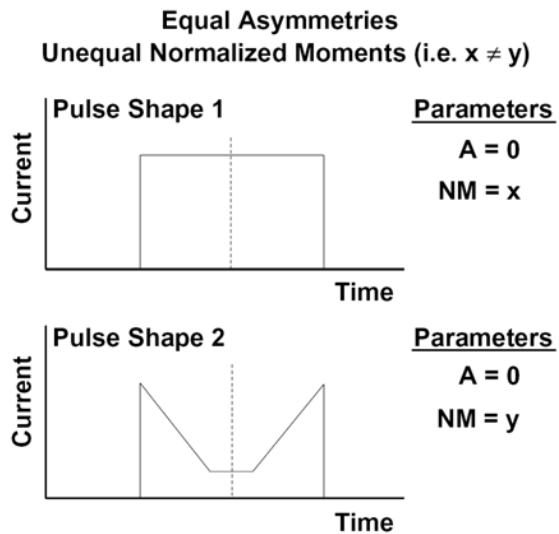

Figure 3: Demonstration of how the normalized moment parameter distinguishes between pulse shapes having equal asymmetry parameter values.

Implementation of the discriminator involves first acquiring calibration data containing pulses from the events or event-type of interest. For example, if one wishes to train the discriminator to select single-site events, it is possible to use events from a double escape peak. In a gamma-ray spectrum, double escape peaks are the result of the process of gamma-ray pair production in the germanium crystal. The pair production process takes place at a single location when a suitably energetic gamma-ray interacts with the electric field of an atom's nucleus. The formation of the spectral peak is due to the special case in which both of the two 511 keV annihilation photons escape the crystal without further interaction. Thus double escape events are lower in energy than the initiating gamma-ray's energy by precisely twice the electron mass (i.e. 1022 keV). In this study the calibration pulses were selected from the double escape peak (energy range 1590.9 keV to 1595.0 keV) produced by the 2614.5 keV gamma ray of [208]Tl. The lower bound of the selected energy range was chosen to be the trough location between the [228]Ac 1588.2 keV peak and the 1592.5 keV double escape peak. The upper bound was chosen such that the high energy tail of the 1592.5 keV peak was approximately an equal contribution as the continuum at the upper bound energy. This restricted energy range used to select double escape peak events is estimated to produce a calibration data set composed of approximately 56% double escape events, while the remaining events are either part of the high energy side of the [228]Ac full energy deposition peak or the Compton continuum which is produced by the 2614.5 keV gamma ray of [208]Tl.

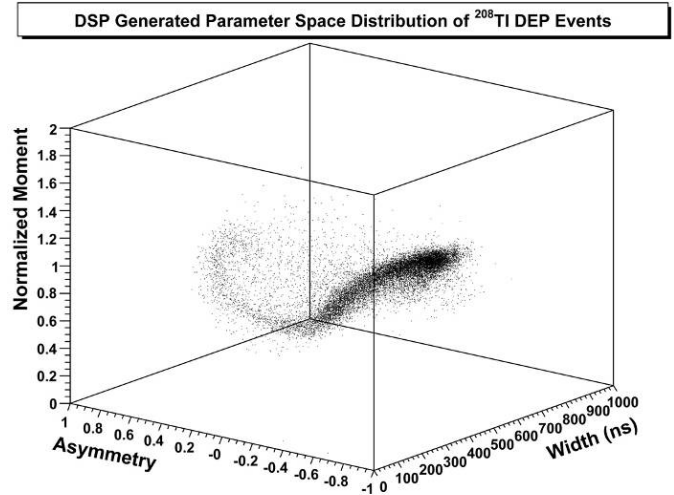

Figure 4: The 3-dimensional parameter-space distribution of double-escape-peak events. These events were taken from 1592.5-keV DEP from $^{208}$Tl and generated on the DSP.

Having a calibration data set now predominately composed of pulses from event-types of interest, each event is located in the three-dimensional parameter-space by the values of the pulse-shape parameters calculated for that event. Presumably, because of the above described event selection process, the bins of the three-dimensional parameter-space having the greatest number of event counts (i.e. density) represent ranges of pulse-shape parameter values *indicative* of the types of events selected for the calibration dataset. To build a quantitative discriminator, the bins of the three-dimensional histogram generated by this method are sorted by descending number of counts per bin. The acceptance region of the discriminator is set based upon selecting all those bins with the highest counts and whose combined total number of counts is a pre-defined fraction of the total events in the calibration data set[2]. For example, if the pre-chosen percentage is 85% of the total number of events, then all bins having the highest event count and whose combined total number of counts is less than or equal to 85% of the total, are included in the discriminator's acceptance region. Thus, in the three-dimensional parameter-space formed by the pulse shape parameters, those bins composing the acceptance region of the discriminator are deemed to have a higher probably for containing events of interest. The pre-chosen percentage, in this example 85%, is the user's handle for increasing the degree of discrimination at the expensive of total signal counts. A figure of merit equation, as presented in [12], is a means to study this handle, but was not part of the present work. Figure 4 shows the three-dimensional parameter space histogram for a calibration data set composed of events under the double-escape peak of the 2615 keV gamma ray of $^{208}$Tl.

---

[2] Although not the scope of this paper, one of the authors has studied the effect of bin size on this histogram-based method. It suffices to say, a "reasonable" bin size producing a broad distribution of event counts per bin is needed. It is possible to make this statement quantitative at the expense of computational speed; however, only slight differences in the results are seen over a broad range of "reasonable" bin sizes.

## 3. Digital Signal Processor Implementation

The hardware used for data collection and processing is a product of X-Ray Instrument Associates (XIA). Pulses generated in the germanium move through an integrating charge-sensitive preamplifier and are sampled as voltage values in XIA's DGF Pixie-4 acquisition card at 75 MHz with 14-bit ADCs. A field-programmable gate array (FPGA) is used for trigger trapezoidal filtering and pileup inspection as chosen by the user. On a confirmed trigger, a digital signal processor (DSP) reconstructs the pulse height and may perform additional tasks depending on the application. Figure 5 shows a block diagram of the Pixie-4 acquisition card.

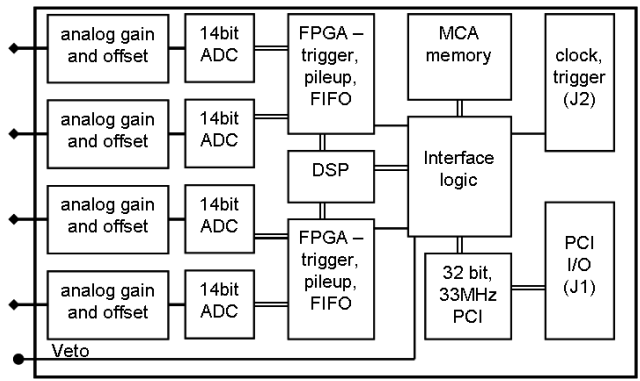

Figure 5: Pixie-4 spectrometer hardware block diagram [16].

Digital signal processors are known for their ability to process data and perform filtering at very fast rates. The high processing speeds are in large part due to the multiply-accumulator (MAC) that can multiply numbers and accumulate with very few instructions. They are optimized for matrix operations such as convolution and dot products.

The Pixie-4 acquisition card uses a 16-bit fixed-point DSP capable of executing 33 MIPS. The DSP is a 2185 from Analog Devices. The computational units located on this DSP are: one Arithmetic Logic Unit (ALU), one Multiplier/Accumulator (MAC), and one Barrel Shifter. The ALU takes two 16-bit inputs and provides a 16-bit output. The MAC takes two 16-bit inputs for multiplication and has a 40-bit output that consists of two 16-bit and one 8-bit register. This output can be fed back to the input of the MAC allowing for accumulation. The Barrel Shifter has a 32-bit output, consisting of two 16-bit registers, of the shifted data. Barrel shifters differ from normal shifters in digital electronics in that they allow multiple bits to be altered in one clock cycle.

A feature available to users of the Pixie-4 is the ability to write DSP code and compile it into the firmware for custom pulse processing. This was the feature used to implement the parametric pulse-shape analysis described in Section 2.

XIA recommends writing all code in assembly as they found compilers (e.g., C compilers) inflate the code causing slow system execution. Writing directly in assembly also allows users to keep code concise and efficient. For this reason, familiarity with the ADSP2100 family instruction set is useful to develop custom DSP code for the Pixie-4. XIA provides a pointer to the start of the pulse data, the length, and energy values, among other variables. This is necessary for traversing the pulse data during filtering operations.

In the ADSP2185, one way to move data back and forth is through the data and program memory bus. Data memory has 16-bit registers and program memory contains 24-bit registers.

However, one constraint in writing custom DSP code for XIA hardware is that only data memory may be used for data storage and transfers.

The DSP uses fixed-point integer math for processing known as 16.0 format. In this format, the radix point falls at the end of the number. This means values must be scaled to fall within $-2^{15}$ and $2^{15}-1$ for signed values and 0 and $2^{16}$ for unsigned values. Two's compliment is used for the representation. So a number in 1.15 format would need to be multiplied by $2^{16}$ to scale it to 16.0 format.

This conversion was important for coefficient scaling when implementing the initial part of the discriminator where Savitzky-Golay (SG) [13] filtering is performed. SG filtering is a type of low-pass filter where a least-square fit to an $n^{th}$ order polynomial is performed on a moving window of a user-defined length. It is generally used to smooth data as it retains most of the spectral features. Another advantage is that it may inherently perform a smoothed $n^{th}$ order derivative, a feature used here to obtain current pulses by taking a first derivative of the charge pulses generated from an integrating preamplifier. There exists a set of pre-computed coefficients that perform the Savitzky-Golay filtering depending on filter requirements [14]. The sum of the coefficients is one and each is in 1.15 format. To use this filter in the DSP, the pre-computed coefficients were shifted by $2^{16}$ to convert them to 16.0 format.

Initially, the original pulse was copied to a temporary buffer and the SG filter was performed on the data in the temporary buffer. A challenge in doing this was the possibility of using previously filtered data points to calculate subsequent points. To avoid this, a temporary circular buffer was allocated that contained the unaltered pulse data pertaining to the front part of the moving window. When a filtered data point was calculated, the unaltered point was copied to the temporary buffer, the pointer incremented, and the original data point then written over. When the filter was applied to the next data point, the pointer in the circular buffer would be pointing to the correct data point. Figure 6 illustrates this process. The blocks $S_n$ are smoothed data points, $U_n$ are unsmoothed data points, and $C_n$ are the SG coefficients. In this case, the unsmoothed data point $U_7$ was just copied to the circular buffer and the smoothed point $S_7$ was written to the pulse data memory location. To calculate $S_8$, the MAC would use the circular buffer and the five data points starting from the pulse data pointer and take the dot product of those with the SG coefficients. This methodology was used to calculate the SG filter when necessary.[3]

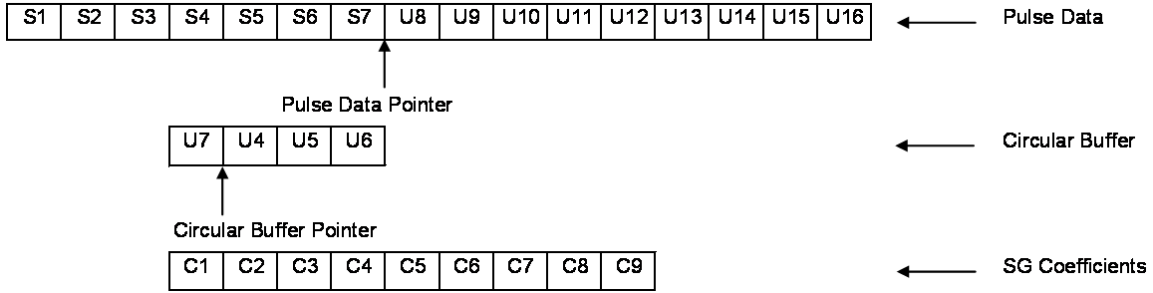

Figure 6: Savitzky-Golay filter implementation

The pulse width was calculated by first finding the maximum value of the pulse. A constant fraction discriminator (CFD) value of 8 percent was used to find the front and back locations of the pulse width. The choice of 8 percent is based upon being safely above the noise of the signal baseline; ideally this CFD is set as low as the noise level allows. The pulse width calculation process involved taking 8 percent of the peak, then traversing the pulse in both directions from the peak until the CFD limit was reached. The address of the start and end locations of the pulse were stored for further parameter calculations. The difference of start and end locations was the pulse width used and stored in a 16-bit channel header register.

The pulse asymmetry calculation involved first finding the midpoint between the start and end address of the pulse. The difference of start and end address was just shifted right by one bit to find the midpoint, effectively dividing by 2. Subsequent to the location of the midpoint, the front and back halves of the pulse were integrated. These values were stored for further processing. Since the possibility that the area for either front or back might exceed 16 bits, 32-bit addition and subtraction was performed. This involved performing addition with carry, and a subtraction with borrow. Each result was stored in two 16-bit registers.  Eq. (1) was used to calculate the pulse asymmetry.

---

[3] This method was originally developed to operate using a minimally sized memory space. Although the technique used was not necessitated in this case, there was no loss in reusing the already developed algorithm.

The ADSP2100 family instruction set provides two instructions for performing a division, DIVS and DIVQ. DIVS is executed first, for signed division, followed by 15 executions of the DIVQ instruction. These instructions implement a conditional add/subtract non-restoring division algorithm. This division algorithm takes a 32-bit dividend and 16-bit divisor outputting a 16-bit quotient. Since the divisor could potentially exceed 16 bits, it is scaled. If it was 16 bits or less, the number was left intact. If it was larger than 16 bits, the most significant bit was shifted to bit 15 of the 16-bit divisor and the shift number was stored. Bit 16 of the divisor served as a sign bit.

The value of the pulse asymmetry parameter should vary between −1 and 1. Since the DSP uses integer math, when the value is read out, the shift value may also be read out. For readability outside the DSP, this parameter should be divided by 2 raised to the shift value to obtain the floating-point representation between −1 and 1. This converts 16.0 format to 1.15.

The normalized moment was calculated by using Eq. (2). The pulse was traversed from start to end locations. The mid-point value of the pulse was loaded and used to subtract the current location ($i$-$N_{mid}$). The MAC has a 16-bit register that is fed back to the input of the multiplier through a multiplexer (MUX) for performing smaller multiplications such as squaring numbers. This feedback register was used to calculate the squared portion, ($i$-$N_{mid}$)$^2$. This value was fed to the input of the MAC along with the current location value, $j_i$. These values were accumulated and the result was multiplied by 12. This provided the dividend portion of the total moment of the pulse and was stored in two 16-bit registers.

The divisor portion of the normalized moment was calculated by taking the total area under the pulse, which was previously found in the pulse asymmetry calculation, and multiplying this by the square of the width. This number potentially exceeded 16 bits, so scaling was necessary. Since the normalized moment is always a positive number, unsigned 16-bit representation was used. This allows another bit for representing the value. To perform the scaling, the most significant bit of the divisor was shifted to bit 16 of the register and the shift amount was stored in a separate location.

The normalized moment falls between 0 and 2. When this value is read out from the DSP, the shift value is also available. In order to scale the number correctly outside the DSP, the output should be divided by 2 raised to the shift value. This converts the number to the floating point representation that falls between 0 and 2.

## 4.  Comparison and example application of PSD parameters

To test the parameters generated with the DSP, a discriminator was constructed to enhance double-escape-peak events of the 2615 keV gamma ray from [208]Tl. Data was taken on a 130% Ortec high-purity germanium detector. The DSP firmware was loaded onto the XIA unit and several sets of data were taken.

The three discriminator parameters were calculated with both the DSP and an offline analysis tool. The double-escape-peak events were located on the energy histogram and used to build the discriminator. The 3-D parameter-space plot for the DSP method is illustrated in Figure 4. The plot for the offline analysis is visually indistinguishable (not shown).

Although both methods of analysis generate similar parameter space distributions, a closer look at the individual parameter distributions reveals slight differences. The differences in the three parameters were calculated (offline value minus DSP value) and histograms were generated to display the differences. Figure 7 illustrates the three difference plots.  In pulse digitization, each sample unit is 13.3 ns in length. The width difference plot represents less than one sample unit of difference, specifically 2.3 ns offset between the DSP calculated width and the offline analysis tool. For all events in the energy range 1575–1605 keV, the root mean square deviation about the mean offset was 6.5 ns. As a detail, because the offline analysis utility interpolates between ADC sampling points to 1 ns precision, the width difference distribution appears continuous rather than in 13.3-ns steps as would otherwise be expected. The result of these details, and the average 2.3-ns greater pulse width calculated by the offline utility, is that the calculated pulse midpoint is displaced from that calculated by the DSP. Through an examination of all 27 possible combinations of 13.3-ns sample step differences (i.e., −1, 0, +1 sample steps) in each of the three left, middle, and right window-point assignments, it was demonstrated such a disagreement on the midpoint assignment will cause the majority of counts in the pulse asymmetry difference histogram to fall in the positive region. These differences in width and area get propagated through the normalized moment equation and again appear as a shift of the distribution away from zero. Finally, we take note of the tails of the distributions exposed by plotting the difference histograms with a log scale. These tails contain less than 1% of the events used in the study and are attributed to a pulse rejection feature included in the offline analysis procedure.

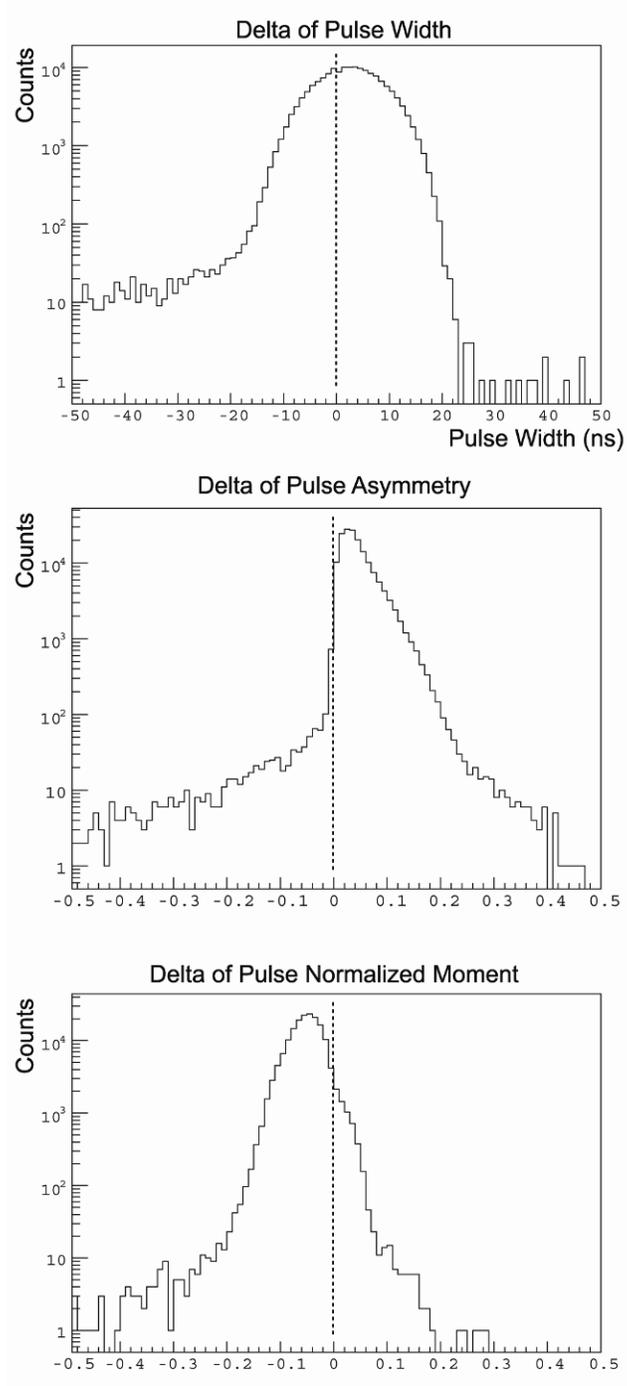

Figure 7: Histograms of the difference between discriminator parameters generated from DSP and the offline analysis.

A significant feature of this pulse-shape discriminator is that it is self-calibrating, that is, it does not require pulse shape simulation input. Recall an initial calibration dataset is collected that is composed of events (or events like those) of interest. In this way, so long as the same parameter calculation method is used on the calibration dataset and the analysis dataset, the discriminator will function correctly in the sense of preferentially selecting events resembling those of the calibration dataset. Differences that exist between the parameters generated from the two methods only highlight the details of translating the original C++-based algorithm into DSP-code. No additional burden is created by the implementation choice (DSP vs. offline) as *any* PSA method must be individually evaluated in a rigorous analysis setting. Figure 8 shows an application of the DSP-based multiplicity discriminator following the same analysis process as presented in [12]. The single-site, double-escape-peak (DEP) events of the 2615-keV gamma ray of $^{208}$Tl are preferentially selected relative to multi-site events, such as those found in full energy peaks.

Table 1

Percentage of events retained by the two implementations of the multiplicity discriminator compared to the true fraction of single-site events predicted by a Monte Carlo simulation.

| | Percentage of Events Retained | | |
| | by the multiplicity discriminator | | as true single-site events |
| Event Type | DSP-based | Offline Analysis | Simulation |
| --- | --- | --- | --- |
| Full Energy Peak 1588.2 keV ($^{228}$Ac) | 48.6 ± 0.5 | 54.9 ± 0.3 | 1.04 ± 0.03 |
| Double Escape Peak 1592.5 keV ($^{208}$Tl) | 101.1 ± 1.2 | 96.8 ± 1.5 | 69.3 ± 0.8 |
| Compton Continuum 1596 - 1600 keV | 62.2 ± 1.1 | 64.7 ± 1.3 | 30.9 ± 1.1 |

Table 1 presents a comparison of the discriminator's performance when implemented in the DSP and offline analysis. The number of events in a peak or continuum region is estimated from the result of a fit to the data using a flat continuum and three independent Gaussian distributions.

The errors reported are based solely on event statistics. In this case the discriminator is set to retain nearly 100% of the events from the double escape peak, which is used as a surrogate for a source of single-site events. Approximately half of the neighboring full energy peak events from the 1588 keV gamma-ray of [228]Ac are rejected, as are approximately one-third of the Compton continuum events in the energy range 1596 - 1600 keV. A simple Monte Carlo simulation of 2614.5 keV and 1588.2 keV gamma-rays was developed to compare the discriminator results to expectations from the underlying physics processes. In the simulation it is possible to select the true fraction of events which are single-site in nature (i.e., no fitting procedure is required). These fractions are also present in Table 1. There are several observations. First, there are apparently non-statistical differences between the DSP and offline discriminators at approximately the 5% level. We take this as a measure of our systematic error. Second, the discriminated fraction of events does not match the expectations from the simulation. One may point out the radial position resolution in a germanium detector is limited and thus two interactions located within, say, 1 mm should not be expected to be disguisable from the pulse shape. To check this possibility, for each event the maximum radial distance between any two interaction locations was calculated from the simulation results. Even when the simulation definition of "single-site" is broadened to include all events with a radial extent less than some value (call it the radial position resolution), no consistent match is found with the data[4]. Third, in fact only 69% of the double escape peak events are single site events. The remaining events contain cases such as a Compton scattering followed by pair production or pair production followed by (reabsorbed) bremsstrahlung radiation. Combining this 69% with the 56% of the events we attribute the double escape peak, our sample population of events used in the calibration step is less than 40% single-site. All of these observations point to the not unexpected need for detailed analysis when this discriminator is employed in a particular application. The scope of the current work is to demonstrate an online DSP-based implementation of the discriminator rather than study the specific details of applying the discriminator.

---

[4] We note, the method used here to define the radial extent of an event does not take into account the relative amount of energy deposited at each location, which certainly affects the pulse shape produced.

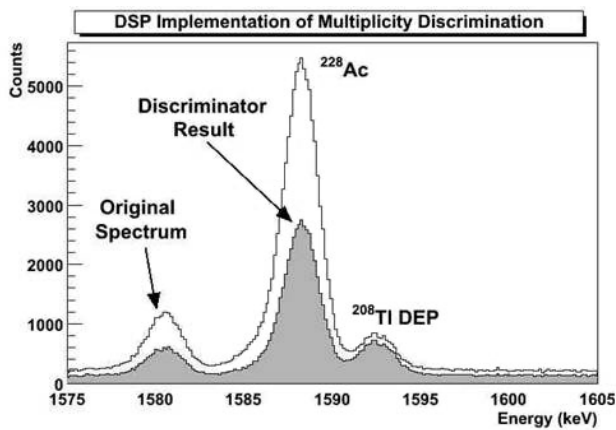

Figure 8: Filtered data using DSP-generated parameters. White is original data and gray area is filtered data showing the selection of double-escape-peak (DEP) events while rejecting a significant fraction of multi-site events in adjacent peaks. The equivalent figure produced by the off-line analysis is, by visual inspection, nearly identical.

To demonstrate the value of online DSP of pulse shapes, several example data runs were collected in series; hence, having identical source and detector configurations. The data acquisition settings were the same from run-to-run except for the run type and whether or not the DSP was turned on. Three run types were considered. The first run type, 0x100, is full-list mode data that records nine 2-byte words and the entire pulse trace for each triggered channel in an event. Run type 0x101 records the nine 2-byte words for each triggered channel in an event, but the pulse trace is not saved. Run type 0x103 is a fast-list mode recording *only* two 2-byte words (time and energy) for each triggered channel in an event. In both run types 0x100 and 0x101, the three DSP parameters are stored as part of the 9-word record. For comparison, each run type was also used when the DSP was explicitly turned off. This was to investigate if there is any overhead in using the DSP. In this test, only one channel was connected to an HPGe detector.

Figure 9 presents a schematic interpretation of the layout of data recorded from the data acquisition system. Each buffer represents data recorded on the acquisition board and then transferred as a block to the connected CPU's memory (i.e., hard drive). Thus, a single run is composed of many "spills" of these buffer blocks. As the size of the data buffer on the acquisition card is fixed, but the different run types record different amounts of data per channel, the number of events recorded in a buffer will vary. Thus, in general, run type 0x100 records the least number of events per buffer while run type 0x103 records the most. A result of this is, for a fixed-duration run, more buffers will be transferred for type 0x100 data in comparison to type 0x103 data.

We define a transfer time, $\Delta t_{Transfer}$, as the period between when the last event in a buffer is recorded and the beginning of the next buffer. We wish to study the total transfer time of the various run types to evaluate the utility of the DSP implementation of the pulse-shape analysis. Table 2 presents results from six short runs demonstrating the duty cycles for each of the run types. The columns with an "XIA" label report information from the acquisition software. The column "File Open" represents the duration between when the computer file system reports the binary data file was first created and last modified. The total live time and transfer time were calculated based on independently calculating and summing together the individual live times and transfer times, as presented in Figure 9. The percent transfer time is the total transfer time divided by the sum of the total live time and total transfer time. The average buffer read time is an estimate of the time required to transfer the buffer data block from the acquisition card to the CPU memory and recommence acquiring event triggers. It is calculated as the total transfer time divided by one less than the number of buffer spills.

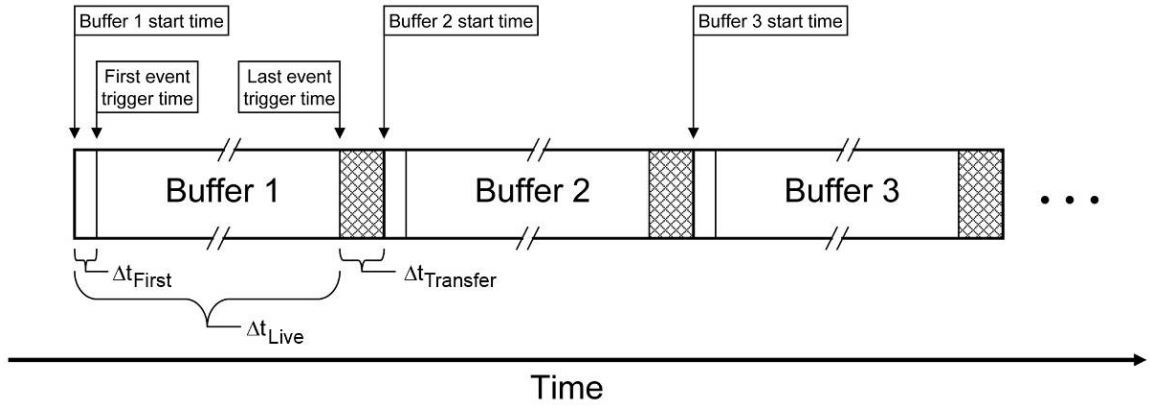

Figure 9: A schematic interpretation of the layout of data recorded to file from the acquisition system.

Table 2 dramatically demonstrates the effect of recording pulse traces – the percentage transfer time is roughly 33%. In contrast, in type 0x101 data when the DSP is calculating the pulse-shape analysis parameters, the transfer time percentage is better by a factor of 20. In comparison to data acquired when the DSP is not turned on, the acquisition card appears to perform essentially the same, thus demonstrating there is little or no overhead in using the on-board DSP. The data also reveals a consistent average value for the time required to transfer the buffer data from the acquisition card to the CPU memory of about 85 milliseconds.

Table 2

Results from six short runs demonstrating the duty cycles for each of the run types

| DSP | Run type | Buffer spills | Events | XIA run time (s) | XIA event rate (Hz) | File open (s) | Total live time (s) | Total transfer time (s) | Percent transfer time (%) | Average buffer transfer time (s) |
|-----|----------|---------------|--------|------------------|---------------------|---------------|---------------------|-------------------------|---------------------------|----------------------------------|
| On  | 0x100 | 1000 | 26000 | 185.6 | 140 | 264 | 177.1 | 86.61 | 32.8 | 0.087 |
| On  | 0x101 | 38   | 25802 | 183.8 | 140 | 187 | 183   | 2.98  | 1.6  | 0.081 |
| On  | 0x103 | 16   | 26176 | 188.3 | 139 | 189 | 187.6 | 1.46  | 0.8  | 0.097 |
| Off | 0x100 | 1000 | 26000 | 186.5 | 139 | 266 | 178.9 | 86.37 | 32.6 | 0.086 |
| Off | 0x101 | 38   | 25802 | 185.1 | 140 | 188 | 183.9 | 3.18  | 1.7  | 0.086 |
| Off | 0x103 | 16   | 26176 | 186.0 | 141 | 187 | 185.3 | 1.22  | 0.7  | 0.081 |

## 5. Conclusion

The multi-parameter pulse-shape discriminator was successfully implemented in a digital signal processor (DSP) within the XIA Pixie-4 multi-channel analyzer. The performance of the multiplicity discriminator suffered no apparent degradation relative to off-line code as a result of the implementation in the DSP. The benefit of this DSP implementation of pulse-shape discrimination is the increased duty cycle of the acquisition hardware. Specifically, it was demonstrated more gamma-ray events were acquired and analyzed with the DSP in real-time than is possible if the pulse waveforms are transferred to external storage for offline pulse-shape analysis. Conversely, the ability to generate the 3-D parameter space for multiplicity discrimination in real-time also removes the need for computationally intensive offline calculations. For example, rather than performing event-by-event pulse shape analysis, a simple restriction to event types of interest can be based on examination of the three parameter values. Thus, these techniques provide a means to broaden the potential applications of pulse-shape analysis in gamma-ray spectroscopy to include field equipment requiring near real-time analysis. Two potential applications for the generalized multiplicity discrimination technique described here are real-time Compton-continuum suppression (rejecting single-site events) and extending field HPGe detectors for fast neutron sensitivity (selecting single-site $^{72}$Ge(n,n'e-) events).

## 6. Acknowledgements

The authors thank XIA for assisting us in our use of their Pixie-4 cards and data acquisition software. We acknowledge Brian T. Schrom for suggesting the comparison of all 27 possible combinations of pulse window left, middle, and right points, as instrumental in our final understanding of the differences seen (Figure 7) between the DSP-generated parameter values and those calculated from the off-line utility. We thank Kay Hass for helping us prepare and submit this work for publication. We thank three anonymous reviewers for suggesting improvements to an earlier version of this work.